\documentstyle[prl,amssymb,amsbsy,amstext,amsfonts,aps,psfig,preprint,floats]{revtex}

\newcommand {\thetaeta}{\theta_{\eta}^*}

\newcommand {\phieta}{\phi_{\eta}^*}
\newcommand {\Gc}{(\mathrm{GeV}/c)^2}

\begin{document}
\tighten
\draft{}
\topmargin -15pt
\title{The $ep \rightarrow e'p \eta$ reaction
at and above the S$_{11}$(1535) baryon resonance}

 \def\uppa{$^{1}$}
 \def\asuaz{$^{2}$}
 \def\cmupa{$^{3}$}
 \def\cuawdc{$^{4}$}
 \def\cnuva{$^{5}$}
 \def\cwm{$^{6}$}
 \def\edinburgh{$^{8}$}
 \def\duke{$^{7}$}
 \def\fsu{$^{10}$}
 \def\fiu{$^{9}$}
 \def\genova{$^{13}$}
 \def\gwudc{$^{11}$}
 \def\ipn{$^{15}$}
 \def\itep{$^{14}$}
 \def\jmuva{$^{16}$}
 \def\knukorea{$^{17}$}
 \def\frascati{$^{12}$}
 \def\mit{$^{18}$}
 \def\nsuva{$^{19}$}
 \def\ohio{$^{20}$}
 \def\oduva{$^{21}$}
 \def\rubltx{$^{23}$}
 \def\rpi{$^{22}$}
 \def\sphn{$^{24}$}
 \def\jlab{$^{25}$}
 \def\ucla{$^{26}$}
 \def\connecticut{$^{27}$}
 \def\umma{$^{28}$}
 \def\unhdurham{$^{29}$}
 \def\urva{$^{30}$}
 \def\usc{$^{31}$}
 \def\utep{$^{32}$}
 \def\uvch{$^{33}$}
 \def\vpsu{$^{34}$}
 \def\yerevan{$^{35}$}

 \author{ R.~Thompson,\uppa\
         S.~Dytman,\uppa\
	K.Y.~Kim,\uppa\
         J.~Mueller,\uppa\
        G.S.~Adams,\rpi\
         M.J.~Amaryan,\yerevan\
         E.~Anciant,\sphn\
         M.~Anghinolfi,\genova\
         B.~Asavapibhop,\umma\
         T.~Auger,\sphn
% ...
         \\(The CLAS Collaboration)
 }


 \address{
 \uppa University of Pittsburgh, Department of Physics and Astronomy, Pittsburgh, PA 15260, USA\\
 \asuaz Arizona State University, Department of Physics and Astronomy, Tempe, AZ 85287, USA\\
 \cmupa Carnegie Mellon University, Department of Physics, Pittsburgh, PA 15213, USA\\
 \cuawdc Catholic University of America, Department of Physics, Washington D.C., 20064, USA\\
 \cnuva Christopher Newport University, Newport News, VA 23606, USA\\
 \cwm College of William and Mary, Department of Physics, Williamsburg, VA 23187, USA\\
 \duke Duke University, Physics Bldg. TUNL, Durham, NC27706, USA\\
 \edinburgh Department of Physics and Astronomy, Edinburgh University, Edinburgh EH9 3JZ, United Kingdom\\
 \fiu Florida International University, Miami, FL 33199, USA\\
 \fsu Florida State University, Department of Physics, Tallahassee, FL 32306, USA\\
 \gwudc George Washington University, Department of Physics, Washington D. C., 20052 USA\\
 \frascati Istituto Nazionale di Fisica Nucleare, Laboratori Nazionali di Frascati, P.O. 13, 00044 Frascati, Italy\\
 \genova Istituto Nazionale di Fisica Nucleare, Sezione di Genova
 e Dipartimento di Fisica dell'Universita, 16146 Genova, Italy\\
 \itep Institute of Theoretical and Experimental Physics, 25 B. Cheremushkinskaya, Moscow, 117259, Russia\\
 \ipn Institut de Physique Nucleaire d'Orsay, IN2P3, BP 1, 91406 Orsay, France\\
 \jmuva James Madison University, Department of Physics, Harrisonburg, VA 22807, USA\\
 \knukorea Kyungpook National University, Department of Physics, Taegu 702-701, South Korea\\
 \mit M.I.T.-Bates Linear Accelerator, Middleton, MA 01949, USA\\
 \nsuva Norfolk State University, Norfolk VA 23504, USA\\
 \ohio Ohio University, Department of Physics, Athens, OH 45701, USA\\
 \oduva Old Dominion University, Department of Physics, Norfolk VA 23529, USA\\
 \rpi Rensselaer Polytechnic Institute, Department of Physics, Troy, NY 12181, USA\\
 \rubltx Rice University, Bonner Lab, Box 1892, Houston, TX 77251\\
 \sphn CEA Saclay, DAPNIA-SPhN, F91191 Gif-sur-Yvette Cedex, France\\
 \jlab Thomas Jefferson National Accelerator Facility, 12000 Jefferson Avenue, Newport News, VA 23606, USA\\
 \ucla University of California at Los Angeles, Department of Physics and Astonomy, Los Angeles, CA 90095-1547, USA\\
 \connecticut University of Connecticut, Physics Department, Storrs, CT 06269, USA\\
 \umma University of Massachusetts, Department of Physics, Amherst, MA 01003, USA\\
 \unhdurham University of New Hampshire, Department of Physics, Durham, NH 03824, USA\\
 \urva University of Richmond, Department of Physics, Richmond, VA 23173, USA\\
 \usc University of South Carolina, Department of Physics, Columbia, SC 29208, USA\\
 \utep University of Texas at El Paso, Department of Physics, El Paso, Texas 79968, USA\\
 \uvch University of Virginia, Department of Physics, Charlottesville, VA 22903, USA\\
 \vpsu Virginia Polytechnic and State University, Department of Physics, Blacksburg, VA 24061, USA\\
 \yerevan Yerevan Physics Institute, 375036 Yerevan, Armenia\\
 }

\date{\today}
\maketitle
\newpage

\begin{abstract}
New cross sections for the reaction $ep \rightarrow ep \eta$ are reported
for total center of mass energy $W$=1.5--1.86 GeV and invariant momentum 
transfer $Q^2$=0.25--1.5 $\Gc$.  This large kinematic range allows 
extraction of important new information about response functions, photocouplings, 
and $\eta N$ coupling strengths of baryon resonances.  Expanded $W$ coverage 
shows sharp structure at $W\sim$ 1.7 GeV; this is shown to come from interference 
between $S$ and $P$ waves and can be interpreted in terms of known resonances.  
Improved values are derived for the photon coupling amplitude for the 
$S_{11}$(1535) resonance.
\end{abstract}







The study of baryon resonances 
is undergoing a significant rebirth
because of new experimental programs at Brookhaven, the Mainz
microtron, the Bonn synchrotron, and Jefferson Lab.  
Continuous, polarized beams and large acceptance detectors are significantly
improving experimental accuracy.  Older studies found a few 
dozen states with a variety of total angular momentum, parity, and 
strangeness~\cite{PDG}.  Modern theoretical work examines the 
microscopic structure in terms of quark and gluon 
interactions.  Empirical constituent quark models (CQM) achieve excellent 
qualitative agreement with a variety of data~\cite{Capstick92} and provide 
important evidence 
that quark excitations in these states are more important than 
gluonic excitations.  Lattice gauge models simulate full QCD; they 
presently calculate moderately accurate values for excited state 
masses~\cite{Leinweber} and show great promise.
These studies require data of much higher quality than 
was previously available.  

Disentangling the wide and overlapping states that populate reaction data 
is an historical problem.  However, reactions involving $\eta N$ final states
couple only to isospin $\frac{1}{2}$ resonances.  Although
$\pi N$ elastic scattering close to $\eta N$ threshold
(total c.m. energy, $W$=1.485 GeV) shows no strong signal
of a resonance, a prominent peak in the total cross section is 
seen for $\eta$ production in $\gamma N$ and $\pi N$ experiments.
This is widely interpreted as the excitation of
a {\it single} resonance, the spin $\frac{1}{2}$, negative parity,
isospin $\frac{1}{2}$ state $S_{11}$(1535)~\cite{PDG}.  ($S$ labels
the $\eta N$ orbital angular momentum.)  
This state has a branching ratio
to $\eta N$ of 30-55\% compared to a few
percent~\cite{PDG,Vrana} for other states.  
These unusual features have encouraged
alternative theoretical efforts to describe the data in terms of a
strong (possibly nonresonant) final state interaction~\cite{Kaiser}.  

Most previous experiments used pion beams.  
An important advantage of electromagnetic experiments is the ability to
extract the matrix elements for $\gamma N \rightarrow N^*$, commonly 
called the photon coupling amplitudes.  These amplitudes are
primarily sensitive to the quark wave function used.  They are
labeled by the $\gamma N$ total helicity and the virtual photon 
polarization and depend on the 
invariant momentum transfer to the resonance ($Q^2$).  For
a spin $\frac{1}{2}$ resonance, there is one transverse 
amplitude ($A_{\frac{1}{2}}$) and one
longitudinal amplitude ($S_{\frac{1}{2}}$).

Photoproduction experiments ($Q^2$=0) have
reaffirmed the strong energy dependence and $S$-wave (isotropic) character 
close to threshold~\cite{Krusche}.  A recent experiment with polarized
photons~\cite{GamPol} has given new values for $\eta N$ decay branching 
ratios of other resonances through interference with the dominant 
$S_{11}$(1535).  

In electroproduction experiments, $Q^2$ is
nonzero and provides additional structure information about the intermediate
state.  Past $\eta$ electroproduction 
experiments~\cite{Brasse,Beck,Breuker,Armstrong} found an unusually 
flat $Q^2$ dependence of $A_{\frac{1}{2}}$ for
the $S_{11}$(1535) in contrast to the nucleon form factors and photon
coupling amplitudes of other established resonances, 
e.g.\ $P_{33}$(1232).  At this time, there is no definitive explanation
for this difference.  Although previous angular distributions
were largely isotropic at all $Q^2$, no detailed
response functions were extracted because of the poor
angular coverage in traditional magnetic spectrometers.  
Here, $\eta$ electroproduction is used to study the 
$S_{11}$(1535) over a broad range of $Q^2$ and $W$.  
At higher $W$ new interference effects are found
that add to our knowledge of $\eta N$ coupling to higher mass 
resonances.  

\begin{figure}[t]
    \psfig{figure=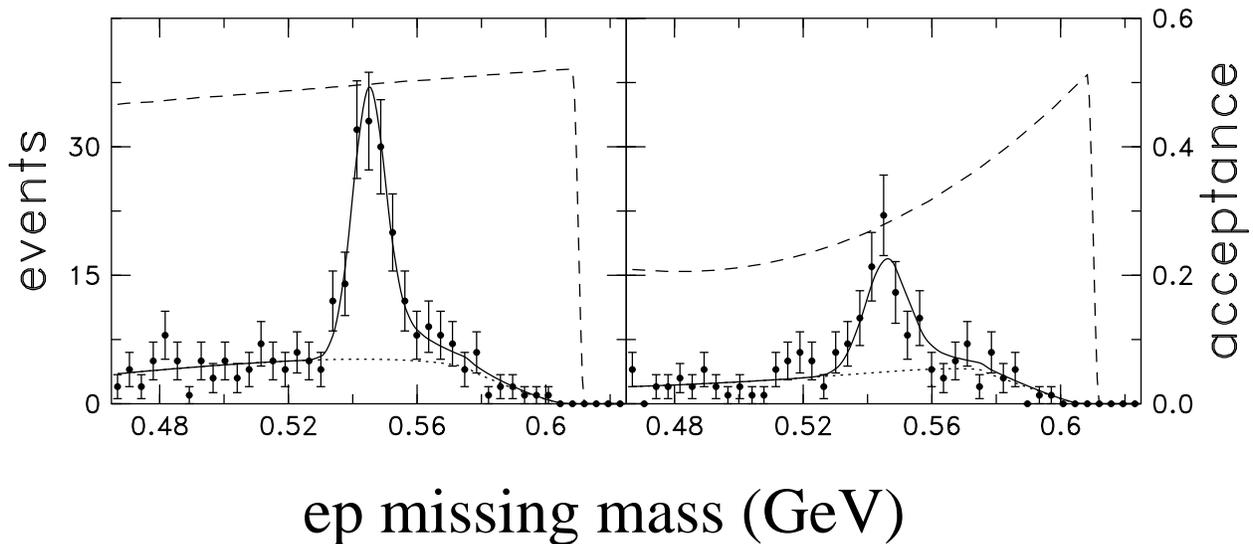,width=\textwidth}
    \caption{Missing mass spectra for $e p \rightarrow e p X$.  Bins shown
are for $W$=1.535 GeV and $Q^2$= 1.25 (GeV/c)$^2$.  Left plot is for
$\phieta$=22.5$^\circ$ and cos$\thetaeta$=-0.4; right plot is for
$\phieta$=67.5$^\circ$ and cos$\thetaeta$=0.0. The dashed line (right scale)
shows the acceptance. The solid and dotted lines are the full fit 
function and the background function only.}
    \label{fg:mm}
\end{figure}

The results reported here used the CEBAF Large
Acceptance Spectrometer (CLAS) at Jefferson Lab (JLab).
It has moderate momentum resolution and excellent solid angle coverage for
final state particles produced in collisions of photon or electron
beams of up to 5.5 GeV energy with various targets.  
This is advantageous for $N^*$ experiments because 
the resonances decay to multiple particles spread over
a large kinematic range.  

The CLAS detector\cite{CLAS} measures angles and momenta of charged
particles for lab polar angles ($\theta$) in the range of
8-142$^\circ$.  
For this measurement, electron beams with energies of 1.645 GeV 
(0.25 $<Q^2<$0.5 $\Gc$) and 2.445 GeV (0.5 $<Q^2<$1.5 $\Gc$) were incident
on a liquid hydrogen target.  A full description of these
results can be found in Ref.~\cite{rat thesis}.
An electron and proton were identified in the final state.  
The hardware trigger identified electrons through threshold
Cerenkov detectors and an electromagnetic calorimeter.
The proton was identified using the time-of-flight technique.
Fiducial cuts were used to restrict particles to detector locations
where the single particle detection efficiency is flat.
Events for the angular 
distributions were binned in $Q^2$, $W$, and the c.m. decay angles 
cos$\thetaeta$ and $\phieta$.  
For the angle integrated cross sections, the same events were binned in $Q^2$ 
and $W$.  

$\eta$ mesons were identified by fitting the missing mass spectrum
(see Fig.~\ref{fg:mm}).
The fit function is the sum of a peak with a radiative tail and a background
function.  The background is due to multi-pion production reactions.  
Lacking a detailed understanding of the background,
a simple function incorporating the proper behavior at the kinematic 
limit and the CLAS acceptance was used.  Acceptance was calculated using a
GEANT-based Monte Carlo simulation of the CLAS detector that included 
bremsstrahlung radiation using the peaking approximation.    
The maximum acceptance of these data is 54\% and no cross section 
is reported where the acceptance was less than 5\%.

\begin{figure}[t]
    \psfig{figure=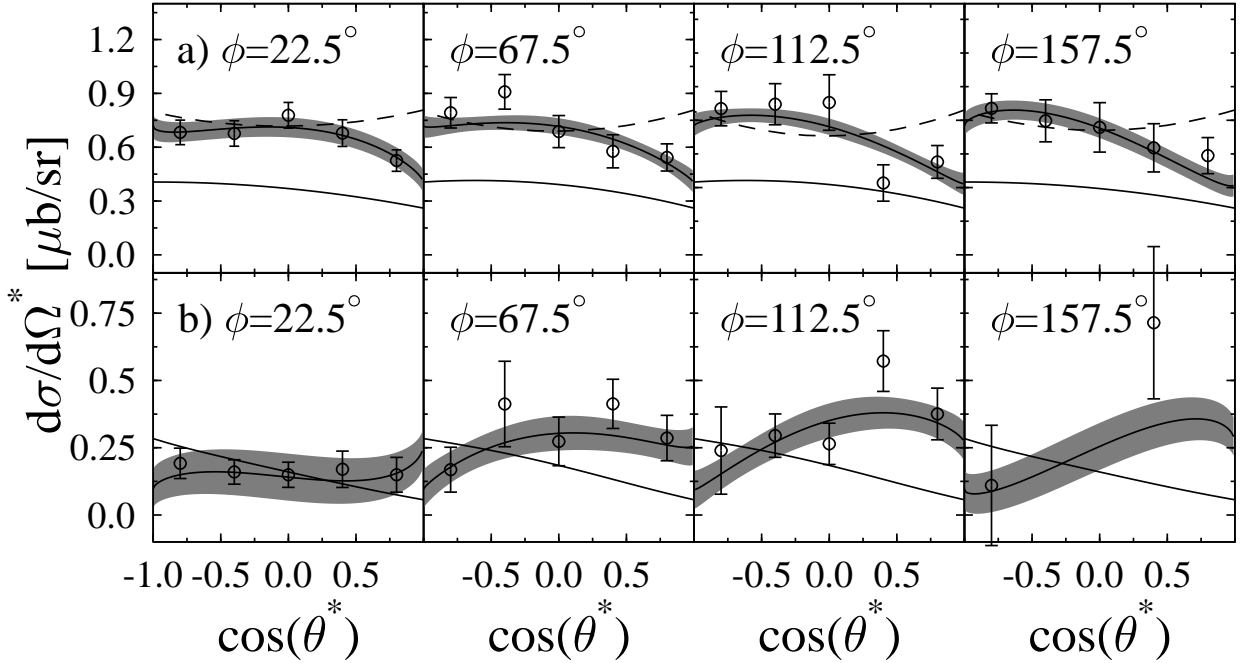,width=\textwidth}
    \caption{Differential cross section for $\gamma_v p\rightarrow p \eta$ 
in the center of mass frame for
a) $W$=1.53 GeV and $Q^2$= 1.25 (GeV/c)$^2$
and b) $W$=1.71 GeV and $Q^2$= 0.75 (GeV/c)$^2$.  Values for $\phieta$ symmetric
about 180$^\circ$ have been averaged.  (No
information is lost this way; see Eq. (1).)  Solid lines with an error band
correspond to the response function fit described in the text.  
Dashed (solid) lines correspond to the effective Lagrangian 
calculation of the RPI (Mainz) group.  See text for details.}
    \label{fg:angdif}
\end{figure}

A detailed study of potential sources of systematic error was made.  
The $ep$ elastic cross section was determined from the same data
set used for the new results.  Agreement
within about 5\% of previous values was obtained, verifying the
efficiency of the hardware trigger to the same level.  
Other studies estimated errors due to inexact knowledge of 
the peak shape and background, 
residual misalignment of the detectors, dependence of the
acceptance on the Monte Carlo input distribution, and variations of the 
fiducial cut edges for the $e$ and $p$. The values varied between 0 and 
10\%.  The total angle-dependent systematic error for each bin was the 
sum of all the components added in quadrature.  
Finally, the total error quoted was obtained by adding the error
in the $\eta$ yield, the acceptance error, and the systematic error in 
quadrature.  

Angular distributions were measured as a function of c.m. decay angles 
cos$\thetaeta$
and $\phieta$ for $W$ for central bin values from 1.5 to 1.83 GeV and
for $Q^2$=0.375, 0.75 and 1.25 $\Gc$.   Sample results for the virtual photon 
cross section in the center of mass frame are shown in Fig.~\ref{fg:angdif}.  
Although all distributions have a significant isotropic component,
deviations from isotropy are seen.  

Fig.~\ref{fg:angdif} also shows predictions by the 
Mainz~\cite{Knochlein} and RPI~\cite{BenMuk95} groups.  
Both models use effective Lagrangians with resonant and nonresonant
terms.  RPI fits parameters to photoproduction
and high $Q^2$ data~\cite{Armstrong}.  Their results are fairly
consistent with the new data at low $W$ but have the wrong slope at 1.62 GeV, 
perhaps due to problems with the u-channel.  The Mainz model 
fits photoproduction data
and extends to finite $Q^2$ using a CQM.  They match the new data
better at high $W$ than RPI, but have the wrong magnitude at low $W$.

\begin{figure}[t]
    \psfig{figure=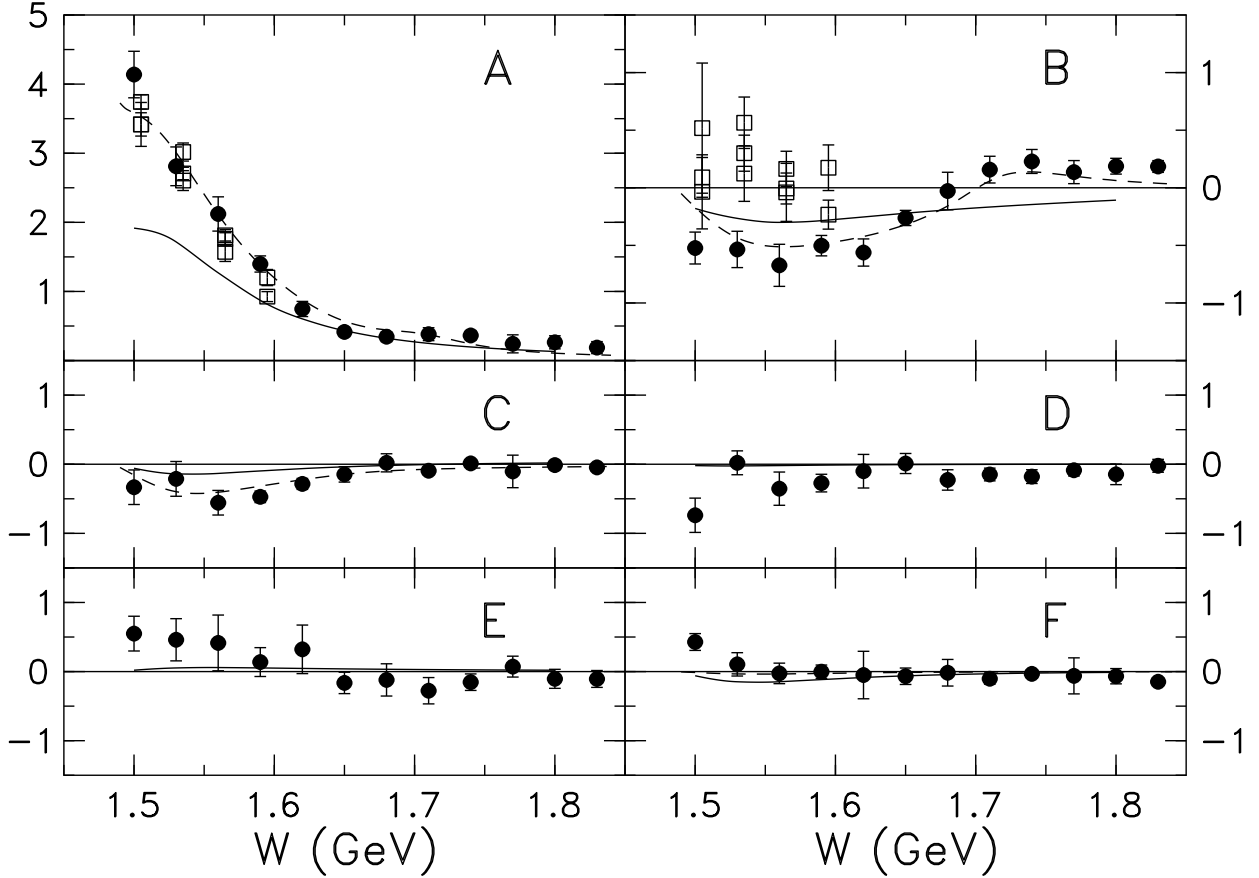,width=\textwidth}
    \caption{Results of fitting the $Q^2$=0.75 $\Gc$ angular 
distribution data of this experiment to Eq. (2).  
See text for details.  Open squares are previous data [8]. 
Contributions from both statistical and systematic sources are displayed.
The solid line is the theoretical prediction of the Mainz group and
the dashed line is a five resonance fit to $A$, $B$, $C$, and $F$.}
    \label{fg:parms}
\end{figure}

The exact virtual photon cross section is 
\begin{eqnarray}
\frac{d^{2}\sigma}{d\Omega_{\eta}^{*}} 
   &=& \frac{|p_{\eta}^{*}|}{K_{cm}}  \left[ R_T + \epsilon R_L  
+ R_{LT} \cdot \sqrt{\frac{\epsilon}{2} (\epsilon +1)} \cos\phieta  \right. \nonumber\\ 
   & & \left. + R_{TT} \cdot \epsilon \cos 2\phieta \right]  . 
\end{eqnarray}

The angular distributions were fit to a form,

\begin{eqnarray} 
   &\approx&  \frac{|p_{\eta}^{*}|}{K_{cm}} 
    \left[ A + B\cdot \cos \thetaeta + C\cdot  P_2(\cos \thetaeta) \right. \nonumber\\ 
    & &   + (D\cdot \sin\thetaeta  
    + E\cdot \sin\thetaeta  \cdot\cos\thetaeta)\ \cdot \cos\phieta \nonumber\\ 
     & & \left. + F\cdot \sin^{2}\thetaeta \cdot \cos2\phieta \right] , 
\end{eqnarray}
assuming dominance of the $S_{11}$ partial wave and truncation to 
total angular momentum up to $\frac{3}{2}$~\cite{Knochlein}.  
Here $\epsilon$ is the polarization parameter and $K_{cm}$ is the
equivalent c.m. photon momentum.  The parameters $A$, $B$, and $C$
contain contributions from the longitudinal ($R_L$) and transverse ($R_T$) 
response functions, $D$ and $E$ parameterize the 
longitudinal-transverse interference response function ($R_{LT}$), 
and $F$ contains the transverse-transverse
interference response function ($R_{TT}$).  
Krusche et al.~\cite{Krusche} used a similar form for
photoproduction data, but only $R_T$ contributes there.
Fit results for $Q^2$=0.75 $\Gc$ are shown in
Fig.~\ref{fg:parms} along with the Mainz predictions\cite{Knochlein}.  

\begin{figure}[htb]
    \psfig{figure=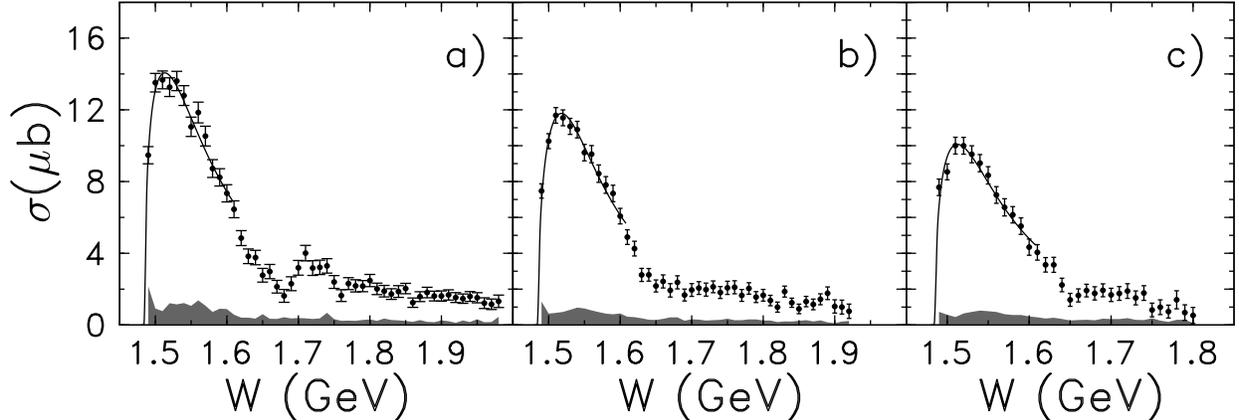,width=\textwidth}
    \caption{New integrated cross section data at (a) $Q^2$=0.625 $\Gc$,
b) $Q^2$=0.875 $\Gc$, and c) $Q^2$=1.125 $\Gc$).  
The shaded band shows systematic errors.
The curves correspond to single-resonance Breit-Wigner fits 
with an energy-dependent width over the energy range shown.}
    \label{fg:intdif}
\end{figure}

Because of their limited
kinematic ranges, previous experiments fit
$A$ and $B$~\cite{Brasse} or $A$ only~\cite{Beck,Breuker,Armstrong}.  
$A$ is mostly due to $S_{11}$(1535) and
is the largest amplitude.  It is probably dominated by the transverse 
amplitude because the longitudinal contribution 
is known to be minor~\cite{Breuker}.  
The $W$ dependence of both $A$ and $C$ are
similar to what was seen in Krusche et al.~\cite{Krusche}.  

Nonzero values for parameters $B-F$ are evidence for interference 
with overlapping resonances or nonresonant
mechanisms.  For the
assumptions made, $B$ and $D$ come from interference between $S_{11}$
and $P_{11}$ partial waves, and $C$, $E$, and $F$ come from
interference between $S_{11}$ and $D_{13}$ partial waves.  
More complete partial wave analyses will be required to disentangle these
contributions in detail.

Limitations in $W$ for previous data are most evident in
$B$.  The value for electroproduction data~\cite{Brasse} was positive 
and poorly determined.  In the photoproduction experiment, $B$ was 
slightly negative.  The new result extends to much higher $W$.
The sign change in $B$ at $W\sim$ 1.7 GeV has not been seen before. 
Such a rapid change is likely due to resonance effects, perhaps  
the onset of $P_{11}$(1710).  In fact, the $W$ dependence of $A$, $B$, $C$,
and $F$ can be reproduced (see Fig.~\ref{fg:parms}) in a simple
isobar model by including $S_{11}$(1535), 
$S_{11}$(1650), $D_{13}$(1520), $P_{11}$(1440), and $P_{11}$(1710) states 
with standard masses and widths~\cite{PDG}.  Neither calculation 
in Fig.~\ref{fg:angdif} includes the $P_{11}$(1710).

$R_{LT}$ and $R_{TT}$ are small compared to the dominant transverse amplitude.
$D$, $E$, and $F$ are consistent with zero over almost all of the range of
$Q^2$ and $W$ covered.  
The Mainz predictions for these amplitudes are small.  Due to a lack of
statistics, the present data provide only a qualitative test of
the predictions.

Angle integrated cross sections were also obtained for events in a given 
$Q^2$, $W$ bin using the same methods as for the angular distributions.  
Distributions in $W$ for three $Q^2$ values  
are shown in Fig.~\ref{fg:intdif}.  Structure is seen at $W \sim$ 1.7 GeV.  
A dip followed by a peak is seen at $Q^2$=0.625 $\Gc$ while
a significant change in slope is seen at other $Q^2$.  This $W$ is where 
$B$ (see above) changes sign and both probably have the same cause.

\begin{figure}[htb]
    \psfig{figure=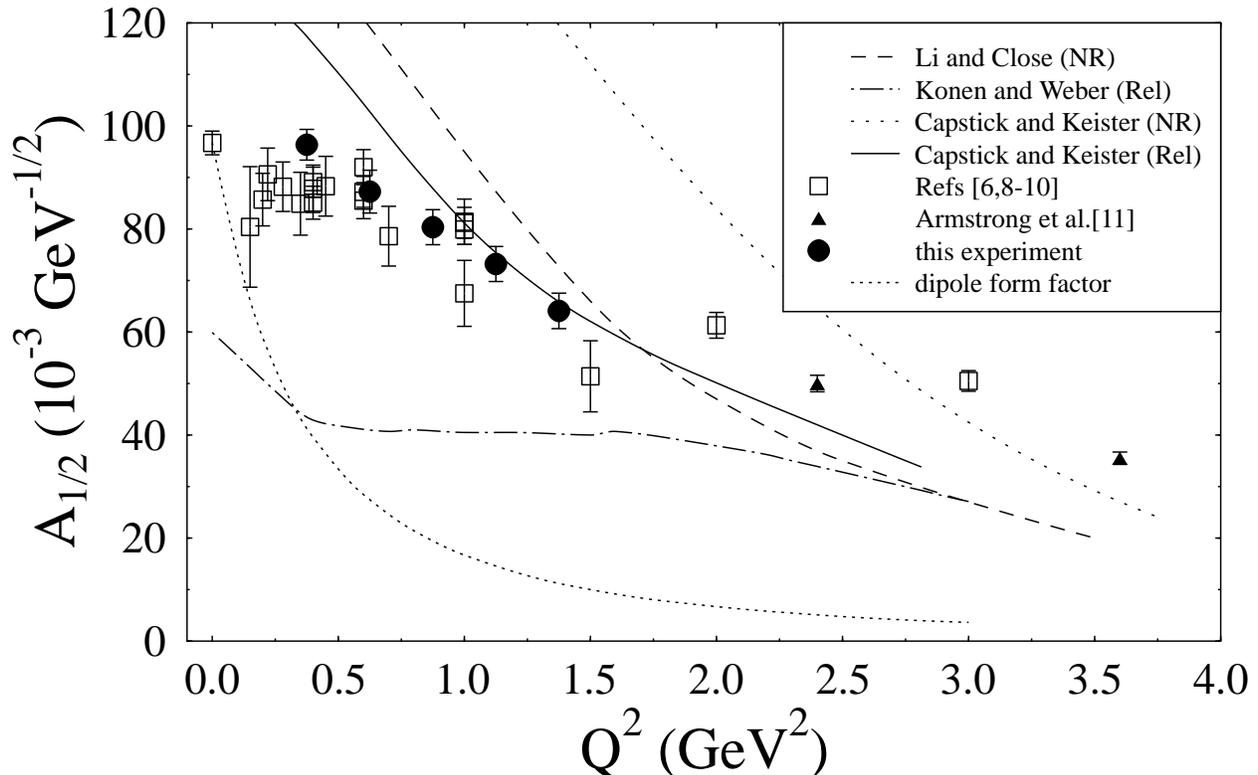,width=\textwidth}
    \caption{Values of the photon coupling amplitude, $A_{1/2}$,
for $\gamma p \rightarrow S_{11}$(1535) obtained from the 
integrated cross section data of this experiment
compared to previous data and various CQM calculations[16].  
All statistical and systematic errors from data 
are included.}
\label{fg:a12}
\end{figure}

Fits to a single Breit-Wigner ($S_{11}$(1535) only) shape~\cite{Knochlein}
are also shown in Fig.~\ref{fg:intdif}.
The nonresonant contribution is ignored in this 
fit~\cite{Knochlein,BenMuk95}.  
Results are very dependent on the $W$ range chosen because other
contributions become prominent at higher $W$.  We find the best fit
with a maximum $W$ of 1.62 GeV.  
These fits give a resonance mass of 1522$\pm$11 MeV
and full width of 143$\pm$18 MeV.  A coupled-channels analysis will
be required to get the most reliable values.  
The maximum cross section for the new and all previous experiments was 
then used to 
determine $A_{\frac{1}{2}}$.  For consistent comparison, 
a full width
of 150 MeV and an $S_{11}\rightarrow \eta N$ branching fraction of
0.55~\cite{Vrana,Armstrong} were used.  
New and re-analyzed old measurements of $A_{\frac{1}{2}}$
are shown in Fig.~\ref{fg:a12}.  A number of values
for $A_{\frac{1}{2}}$ from $\pi$ and $\eta$ photoproduction ($Q^2$=0) data 
with significant model dependence have been reported.  The PDG 
value~\cite{PDG} of 0.09$\pm$0.03 $\mathrm{GeV}^{-\frac{1}{2}}$ reflects this
uncertainty.  The theoretical calculations shown in Fig.~\ref{fg:a12} are
nonrelativistic and relativistic CQM predictions~\cite{CQM}.
None agrees well with the data or each other, an
important failure of the CQM.

These and
other recent data from JLab~\cite{Armstrong} have comparable
values for the $S_{11}$ Breit-Wigner width  
($\sim$ 150 MeV) for $Q^2$ between 0.375 and 3.6 $\Gc$.  
Photoproduction~\cite{Krusche} and Brasse electroproduction 
values~\cite{Brasse} are surprisingly different, 239 and $\sim$90 MeV, 
respectively, in our fits.
At high $Q^2$, the Brasse data is incompatible with 
Armstrong~\cite{Armstrong}; at low $Q^2$, either the single 
resonance interpretation is incorrect or there is a significant
change in dynamics with increasing $Q^2$.

The eta electroproduction data shown here comprise one of the first
results of the CLAS.  It covers the region in $W$ 
at and above the $S_{11}$(1535) resonance 
in great detail.  A consistent
picture of the reaction is given over a more extensive kinematic
range than any previous data.
The values for the interference response functions are small
compared to the dominant transverse response function and
in qualitative agreement with theoretical predictions~\cite{Knochlein}.  
$B$ measures the interference between $S_{11}$ and
$P_{11}$ partial waves.  It is more negative than the photoproduction
results~\cite{Krusche} at low $W$ and changes sign at $W\sim$ 1.7 GeV, 
likely signalling the onset of a strong $P$-wave process.  At the same $W$, 
a sharp change in slope is seen in the integrated data.  
The angular distribution data can be described by a simple 
isobar model using known states.  The $\eta N$ coupling strengths
for these states are poorly known, but can be determined using
these data.
A simple determination of the $\gamma p \rightarrow S_{11}$(1535)
photon coupling amplitude provides new and far more 
consistent evidence for its unusually slow falloff with increasing
$Q^2$.  These data along with other recent Jefferson Lab data~\cite{Armstrong}
provide inescapable constraints on models attempting to describe
the structure of $S_{11}$(1535). 

 We acknowledge the efforts of the staff of 
the Accelerator and the Physics Divisions at JLab that made this experiment 
possible. This work was supported in part by the U.S. Department of Energy 
and National Science Foundation, the French Commissariat 
\`a l'Energie Atomique, the Italian Istituto Nazionale di Fisica Nucleare, 
and the Korea Science and Engineering Foundation.


\begin{thebibliography} {99}

\bibitem{PDG} D.E.~Groom, et al., Eur.\ Phys.\ J.\  {\bf C15}, 1 (2000).

\bibitem{Capstick92} S. Capstick, Phys. Rev. D{\bf 46}, 2864 (1992).

\bibitem{Leinweber} D.B.~Leinweber, et al.,
Phys. Rev. D{\bf 61}:074502 (2000); S. Sasaki, hep-ph/0004252.

\bibitem{Vrana} T.P.~Vrana, S.A.~Dytman, and T.-S.~H.~Lee, Phys. 
Repts. {\bf 328}, 181 (2000).

\bibitem{Kaiser}
N.~Kaiser, T.~Waas and W.~Weise, Nucl.\ Phys.\  {\bf A612}, 297 (1997).

\bibitem{Krusche} B.~Krusche et al., Phys.\ Rev.\ Lett.\  {\bf 74}, 
3736 (1995).  

\bibitem{GamPol} J. Ajaka et al., Phys. Rev. Lett. {\bf 81}, 1797 (1998).

\bibitem{Brasse}
F.~W.~Brasse et al., Nucl.\ Phys.\  {\bf B139}, 37 (1978); F.~W.~Brasse 
et al., Z.\ Phys.\  {\bf C22}, 33 (1984). 

\bibitem{Beck}
U. Beck  et al., Phys. Lett. {\bf B51}, 103 (1974).

\bibitem{Breuker} H.~Breuker et al., Phys. Lett. {\bf B74}, 409 (1978).

\bibitem{Armstrong}C.~S.~Armstrong et al., Phys.\ Rev.\  {\bf D60}:052004 (1999).

\bibitem{CLAS} W.~Brooks, Nucl. Phys. {\bf A663-664}, 1077 (2000).

\bibitem{rat thesis} Richard A. Thompson, Ph. D. Thesis, University of 
Pittsburgh (unpublished), 2000.

\bibitem{Knochlein}
G.~Knochlein, D.~Drechsel and L.~Tiator, Z.\ Phys.\  {\bf A352}, 327 (1995); 
L.~Tiator, C.~Bennhold, and S.S.~Kamalov, Nucl. Phys. 
{\bf A580}, 455 (1994).

\bibitem{BenMuk95} M.~Benmerrouche, N.C.~Mukhopadhyay, and J.F.~Zhang,
Phys. Rev. D{\bf 51}, 3237 (1995); R.~Davidson, priv. comm.

\bibitem{CQM} S. Capstick and B.D. Keister, Phys. Rev. {\bf D51}, 3598
(1995); W. Konen and H.J. Weber, Phys. Rev. {\bf D41}, 2241 (1990); 
F.E. Close and Z. Li, Phys. Rev. {\bf D42}, 2194 (1990) .

\end{thebibliography}
\end{document}